\documentclass[
superscriptaddress,
 reprint,
amsmath,amssymb,
aps,
prx,
]{revtex4-1}

\usepackage{graphicx}
\usepackage{dcolumn}
\usepackage{bm}
\usepackage{hyperref}

\begin{document}

\title{Spatial bunching of same-charge polarization singularities\\in two-dimensional random vector waves}

\author{L. De Angelis}
\affiliation{%
	Kavli Institute of Nanoscience, Delft University of Technology, 2600 GA, Delft, The Netherlands
}%
\author{F. Alpeggiani}
\affiliation{%
	Kavli Institute of Nanoscience, Delft University of Technology, 2600 GA, Delft, The Netherlands
}%
\author{L. Kuipers}
\email{l.kuipers@tudelft.nl}
\affiliation{%
	Kavli Institute of Nanoscience, Delft University of Technology, 2600 GA, Delft, The Netherlands
}%

\date{\today}

\begin{abstract}
Topological singularities are ubiquitous in many areas
of physics. Polarization singularities are locations
at which an aspect of the polarization ellipse of
light becomes undetermined or degenerate.
At C points the orientation of the ellipse
becomes degenerate and light’s electric
field vector describes a perfect circle in time.
In 2D slices of 3D random fields the distribution in
space of the C points is reminiscent of that of
interacting particles.
With near-field experiments we show that when
light becomes truly 2D,
this has severe consequences for the distribution
of C points in space. The most notable change
is that the probability of finding two C points
with the same topological charge at a vanishing
distance is enhanced in a 2D field.
This is an unusual finding for any system
which exhibits topological singularities as
same-charge repulsion is typically observed.
All our experimental findings are supported
with theory and excellent agreement is
found between theory and experiment.
\end{abstract}

\pacs{Valid PACS appear here}
\maketitle

\section{Introduction}
Light-based technology has transformed today's society and will continue
do do so, with applications which range from energy harvesting to
telecommunications and quantum informatics \cite{Hensen2015,Koenderink2015,Polman2016}.
Increasing control over
light's polarization is one key capability
inspiring new developments.
For instance, optical fields near nanostructures
can be engineered to exhibit locations of circular polarization \cite{Burresi2009,deHoog2015,Rotenberg2015},
allowing applications such as spin-dependent
directional coupling \cite{LeFeber2015}, also with
local solid-state spin into optical information conversion \cite{Gong2017}.
Interestingly, points of circular polarization are singularities of
the light field, also known as C points \cite{Berry2004,Gbur2016book}, 
widely studied in structured
light beams \cite{Wang2017,Gbur2016,Fosel2017} and representative
of the transverse
spin momentum of light \cite{Bekshaev2015,Bauer2016,Antognozzi2016,Bliokh2017}

More in general, C points are topological defects
of the vector field which describes light's polarization.
Knowledge and study of
topological defects goes way beyond optics.
Currently, dislocations of the local magnetization known as skirmions
are being intensively investigated \cite{Shibata2018,Reichhardt2016,Fert2017}.
In nematic systems, topological defects continuously attract interest
for their fascinating behavior \cite{Shebdruk2017,Yeomans2014}.
Besides fascination, it was shown how this kind of defects
can even govern the physics of biological system \cite{Saw2017},
and that their spatial arrangement
is representative of intrinsic properties
of the system in which they are found \cite{Doostmohammadi2016}.

Interestingly, also the large ensemble of C points
which naturally arises in random light fields
exhibits an emblematic and rigorous spatial distribution \cite{Berry2001,Dennis2002,Flossmann2005,Flossmann2008},
which resembles that of particles in a simple liquid  and
only scales with the wavelength of the interfering waves \cite{Dennis2002}.
However, a random wave field can be realized
in several
ways~\cite{OHolleran2009,Barkhofen2013,Redding2013,Strudley2014,Shi2015,Pierangeli2017,Bender2017}.
The work so far has concentrated on the investigation
of polarization singularities in two-dimensional (2D) slices
through random three-dimensional (3D) fields in the paraxial limit.
The question now arises how limiting the propagation of light to a truly 2D situation, e.g., by confining it on a flat optical chip, would affect the spatial distribution of its polarization singularities.
In such a case, transverse propagation would set a one-to-one
relation between the wave propagation direction and the
direction of the electric field.
Moreover, this would create correlations between
right-handed and left-handed polarization
that are absent in the three-dimensional fields.

By means of near-field experiments we investigate the spatial distribution of C points
in a planar random light field, and reveal crucial
differences with respect to existing paraxial theory~\cite{Dennis2002}.
We demonstrate that confining light propagation in two
dimensions leads to a large increase in the probability
of finding C points with the same topological charge at close proximity. This is an exotic behavior
for topological singularities, which usually 
exhibit same-charge repulsion.
We relate our experimental
findings to light's handedness and
perfectly describe them with
a new theoretical model
developed for the two-dimensional case.

\section{Experiment and Methods}

\begin{figure*}[t]
	\centering
	\includegraphics[width = \linewidth]{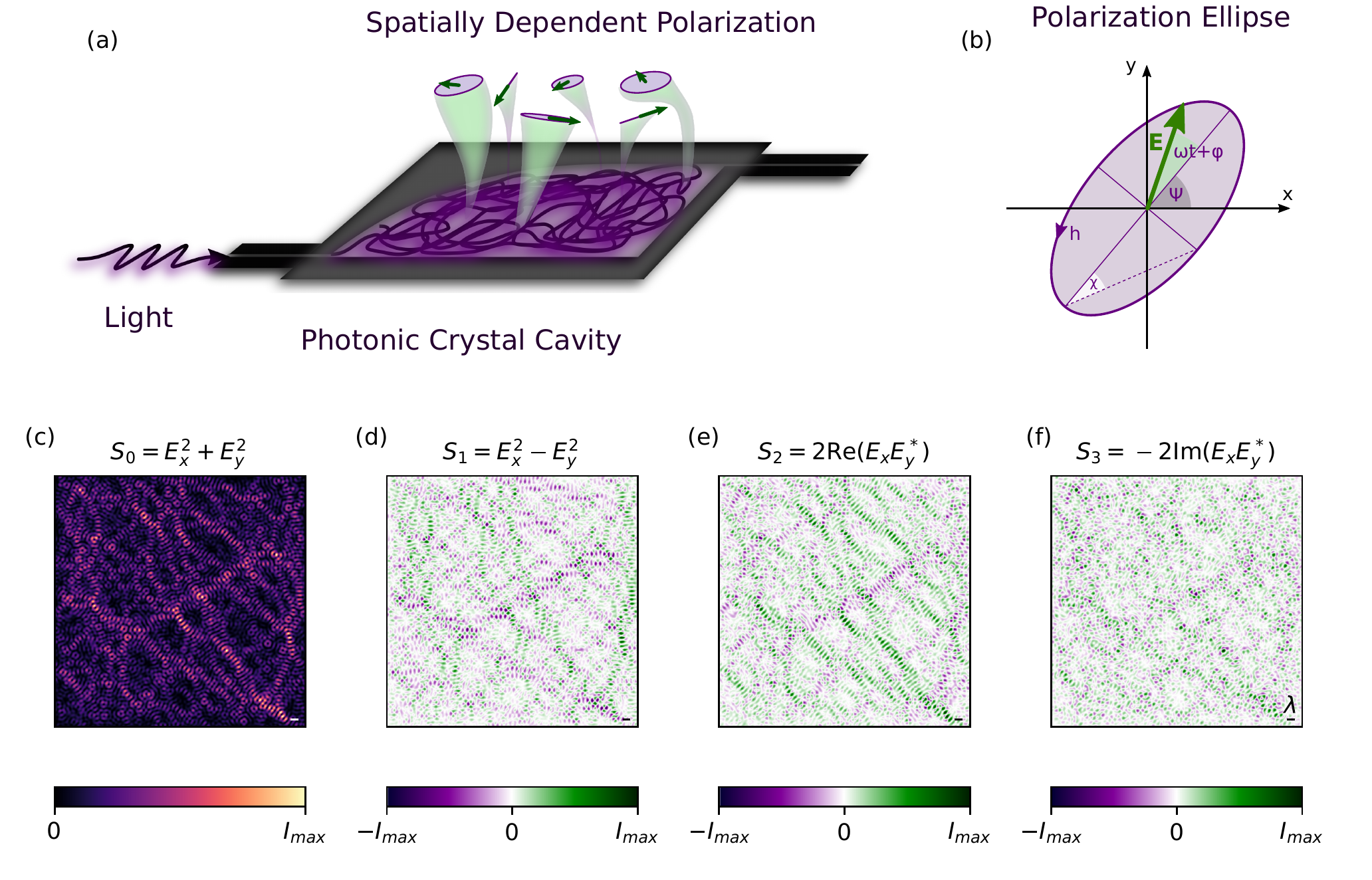}
	\caption{Overview of the near-field measurements of light's polarization in a chaotic cavity.
		(a) Schematic of the experimental realization of random light waves
		in the planar photonic crystal cavity (black). Inside the cavity, light
		exhibits a spatially dependent polarization. This is illustrated by
		a few different polarization ellipses (purple ellipses)
		observed at different points in the cavity (indicated by
		the green shadows).
		(b) Parametrization of the polarization ellipse, describing the local
		polarization state of light. 
		(c-f) Near-field maps of the Stokes parameters of the optical random field
		in a square region of $17\,\mu m\times17\,\mu m$ inside the chaotic cavity.
		$S_0 = E_x^2 + E_y^2$ is the total intensity of the vector field. This is
		displayed with a false-color map ranging from 0 to $I_{max}$, where $I_{max}$
		is the maximum measured intensity.
		$S_i$ ($i=1,2,3$) describe the polarization state of light,
		with respect to linear (horizontal-vertical), linear ($\pm45\deg$)
		and circular (right-left) polarization, respectively. These are also
		represented with false-color maps, which this time range from $-I_{max}$ to $I_{max}$.}
	\label{fig:stokes}
\end{figure*}

\subsection{Near-field Optical Measurements}
In our experiments we map the near field of light waves
propagating in the planar chaotic cavity sketched in Fig.~\ref{fig:stokes}(a).
This is a photonic crystal cavity
realized in a silicon-on-insulator platform (220 nm silicon slab) and
designed to provide random waves propagation~\cite{Liu2015}.
With a monochromatic laser at the telecom frequencies ($\lambda_0\simeq 1550$~nm)
we excite a transverse electric (TE) slab mode
which results in a random superposition of monochromatic TE waves
inside the cavity~\cite{DeAngelis2016,DeAngelis2017}. With a custom-built near-field
scanning optical microscope (NSOM) we probe the light field approximately $20$ nm
above the surface of the cavity.
The measurement of the amplitude and phase of both the in-plane components
($E_x,E_y$)
allows the full characterization
of its polarization state at every point in the measured plane.
For simplicity, we only consider the TE light propagating in the sample,
which has its electric field entirely in the plane of propagation.
We do not investigate TM light, which our cavity was not designed to confine \cite{Liu2015}.

A comprehensive description of light's polarization
is provided by its Stokes parameters \cite{Fowles1989}.
These parameters are often used to characterize the
polarization state of light in the far field,
ranging from a simple laser beam to
the polarized emission
of exotic structures~\cite{Abbas2015}, but they can be
used for a local analysis of the near field as well.
Figures~\ref{fig:stokes}(c)-(f) present the near-field maps of
the Stokes parameters for the optical random field inside the chaotic cavity.
As result of vector light waves randomly interfering, these patterns
are quite difficult to interpret.
However, we can spot a few specific features in the
morphology of each different map.
$S_1$ exhibits patterns of spatial modulation
approximately half a wavelength wide and several wavelengths long.
Depending on their color (sign),
these stripy patterns are either oriented along the $x$ or $y$ axis.
The same observation is valid for $S_2$,
but here the modulations are oriented at $\pm 45\deg$ with
respect to the horizontal axis. No clear preferential direction
stands out from the map of $S_3$.
In fact, $S_1$ is representative of light linearly polarized along
$x$ ($S_1>0$) or $y$ ($S_1<0$), and since light propagates as a transverse vector wave
the observed stripy patterns are reminiscent of $x$-polarized waves mainly propagating
along $y$ and vice versa~\cite{DeAngelis2016}. A totally analogous argument holds for $S_2$,
while $S_3$ doesn't exhibit any pattern that
bears a relation to any specific in-plane direction, being the parameter
representative of circular polarization.

\subsection{Light's Polarization and C points}

A more concise yet comprehensive summary on the complex polarization
pattern illustrated in Fig.~\ref{fig:stokes} can be obtained
from the analysis of its singularities~\cite{Flossmann2008}.
In general, light's polarization is elliptical, thus
parametrized with the orientation $\psi$ of the polarization
ellipse, the ellipticity angle $\chi$ and the handedness $h$ [Fig.~\ref{fig:stokes}(b)].
However, there are special cases in which the polarization ellipse
degenerates into a circle or a line, and some of these parameters
are not well defined anymore.
In two dimensions, such singularities of the vector field
are respectively points of circular polarization (C points) and lines
of linearly polarized light (L lines)~\cite{Nye1983}.

\begin{figure}[t]
	\centering
	\includegraphics[width = 8cm]{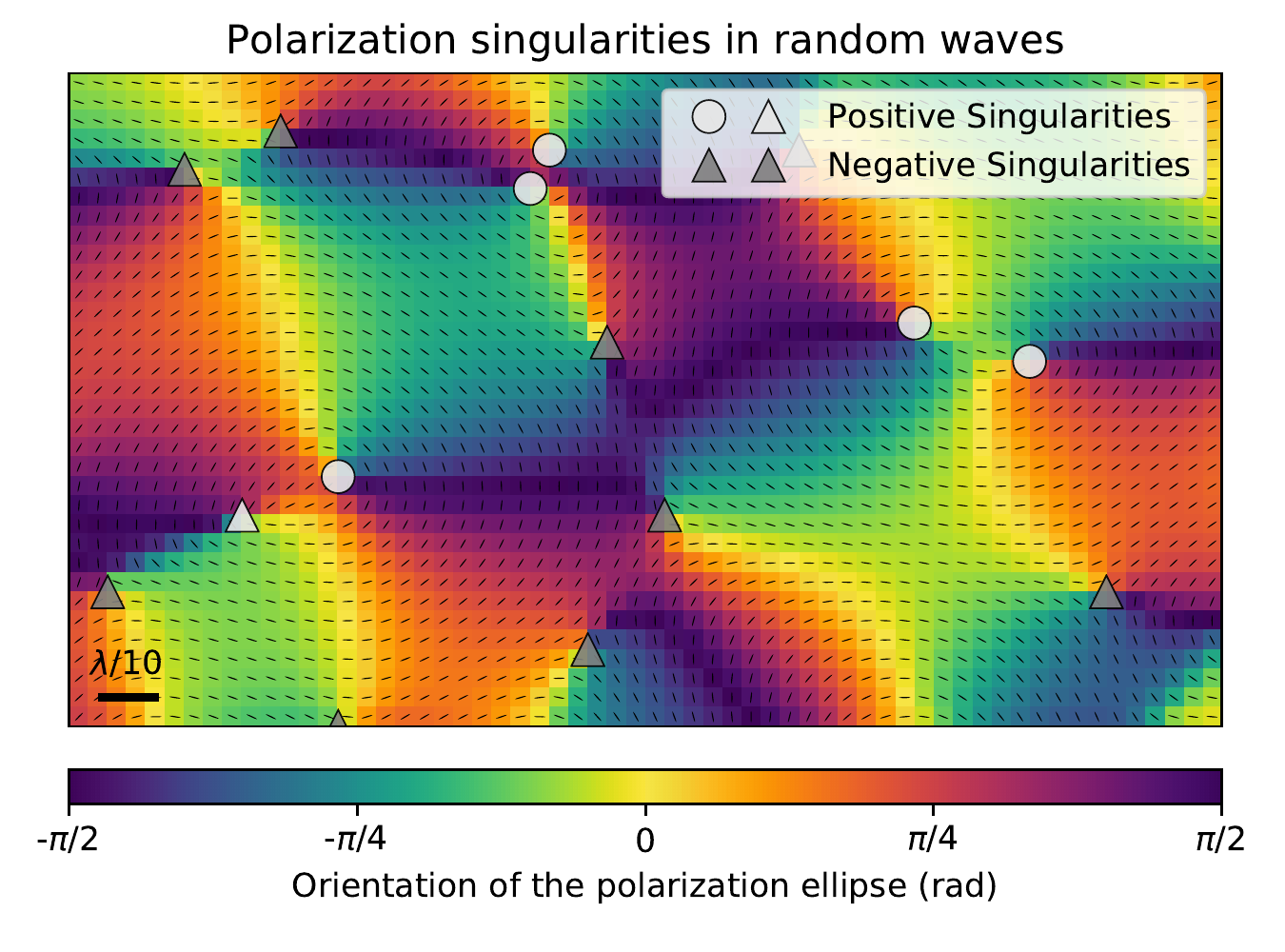}
	\caption{False-color map for the orientation of the major axis of the polarization
		ellipse. The black directors indicate the orientation of such axis too.
		The plot is representative of a subsection of the measured optical random field.
		Circles and triangles are C points. The color of the symbols, white or black, denotes a positive or negative topological charge, respectively. The shape of
		the symbols, triangles or circles, denotes the a star-type or lemon-type
		classification, respectively.}
	\label{fig:polsings}
\end{figure}

Figure~\ref{fig:polsings} is a map of the orientation of the
polarization ellipse for a small subsection of the measurement
presented in Fig.~\ref{fig:stokes}.The
position of C points is highlighted by circles and triangles,
whose color represents their topological charge.
This is defined as the half-integer number of times
that the axis of the polarization ellipse rotates around the singularity,
clockwise (positive charge) or anticlockwise (negative charge).
In Fig.~\ref{fig:polsings}, we only observe topological charges of $\pm 1/2$.

\begin{figure}[t]
	\centering
	\includegraphics[width = 7cm]{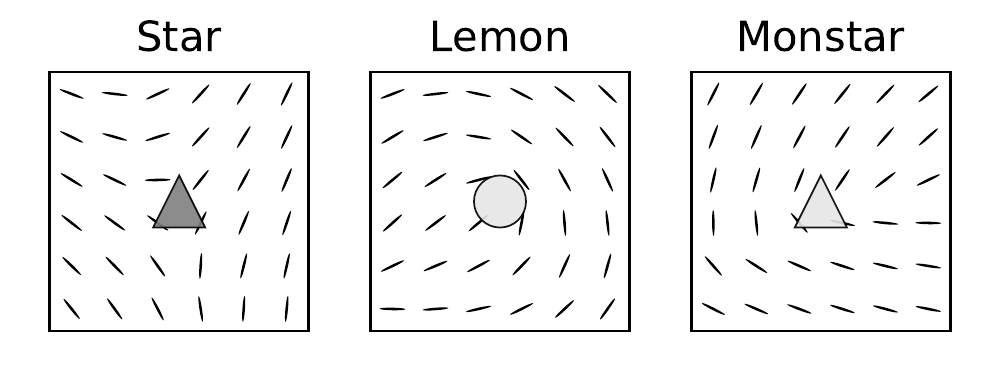}
	\caption{An overview of the three kinds of C points based on their
		line classification~\cite{Dennis2008}. The
		lines are the orientation of the polarization ellipse at
		each pixel around the C-point (circle or triangle),
		as determined from experimental data.}
	\label{fig:lclass}
\end{figure}

Strictly related to their topological charge 
is the so-called \emph{line classification} of C points, which differentiates them in
three types: \emph{lemons}, \emph{stars} and \emph{monstars}~\cite{Dennis2008,Cvarch2017}.
The line classification can be understood by
looking at the orientation of the polarization ellipse
around the singularity, highlighted by the black directors
in Fig.~\ref{fig:polsings} and in the zoomed-in images of Fig.~\ref{fig:lclass}.
For lemon-type singularities (\emph{lemons})
there is only one direction along which
the orientation of the polarization ellipse is directed towards the singularity,
whereas the possible directions are always
three for star-type singularities (\emph{stars} and \emph{monstars}).
To determine the line classification of all the C points
in our dataset in a deterministic
way, we apply the method illustrated
by Dennis for computing the number of directors
pointing towards each singularity \cite{Dennis2008}.
In our figures,
we indicate stars and monstars
with triangles, lemons with circles.

Already a quick glance at Fig.~\ref{fig:polsings} illustrates
the clear relation between topological charge (markers color) and line
classification (markers shape) of C points.
In fact, negative-charge singularities are
always stars, whereas both lemons and monstars are
characterized by a positive charge, as
expected in general for C points \cite{Dennis2008}.
Table~\ref{tab:lclass} lists the fraction of C points for each of the kinds
observed in our experimental dataset.
$50\%$ of the total
number of C points are stars,
and they all carry a negative topological charge.
Approximately $45\%$ of the singularities are lemons,
and only $5\%$ monstars, both types being positively charged.
In the same table, we directly
compare our experimental outcome with
the results from previous
paraxial theory~\cite{Dennis2002} and experiments~\cite{Flossmann2005}.
All these examined statistics
are perfectly consistent with each other.
In summary, the abundance of C points with a particular
line classification is the same for C points in truly
two-dimensional light and two-dimensional slices through
a three-dimensional field.

\begin{table}[h]
	\renewcommand{\arraystretch}{1.3}
	\begin{tabular}{cccc}
		\hline \hline
		Singularity &	2D Field		& \multicolumn{2}{c}{2D Slice of a 3D Field}\\
		Type 		& Experiment		& Experiment \cite{Flossmann2005} & Theory \cite{Dennis2002}\\ \hline
		Star	& $0.4997 \pm 0.0002$	& $0.506 \pm 0.003$	& 0.500		\\	
		Lemon	& $0.4493 \pm 0.0013$	& $0.443 \pm 0.002$ & 0.447 	\\
		Monstar & $0.0503 \pm 0.0013$	& $0.050 \pm 0.003$ & 0.053\\	\hline\hline		
	\end{tabular}
	\caption{Fraction of C points with different line classification.
		The results of our 2D experiment are compared with a previous
		experiments~\cite{Flossmann2005} and theory~\cite{Dennis2002}.}
	\label{tab:lclass}
\end{table}

\section{Spatial Distribution of C points}
\subsection{Pair and Charge Correlation Function}\label{sec:Cdistr}
Having established that there is
no difference between the abundances of the various types
of singularities observed in 2D slices of 3D light fields and
truly 2D fields, the question now arises whether their
distribution in space is also the same.
The natural way of investigating the spatial distribution
of point-like singularities,
is determining their
pair correlation function $g(r)$. Given a C point, this function
describes how the density of the
surrounding C points varies as a function of distance.
This method is widely used to describe
the physics of discrete systems~\cite{Barker1976,Lyubartsev1995,Baldini2016,Frandsen2016,Thaicharoen2017,Sobstyl2018},
it can be directly related to the structure factor \cite{Hansen1990},
and it represents a spatial analogous
of the degree of second-order coherence $g^{(2)}(\tau)$,
commonly used to determine photon bunching and antibunching~\cite{Snijders2016}.

\begin{figure}[t]
	\centering
	\includegraphics[width = \linewidth]{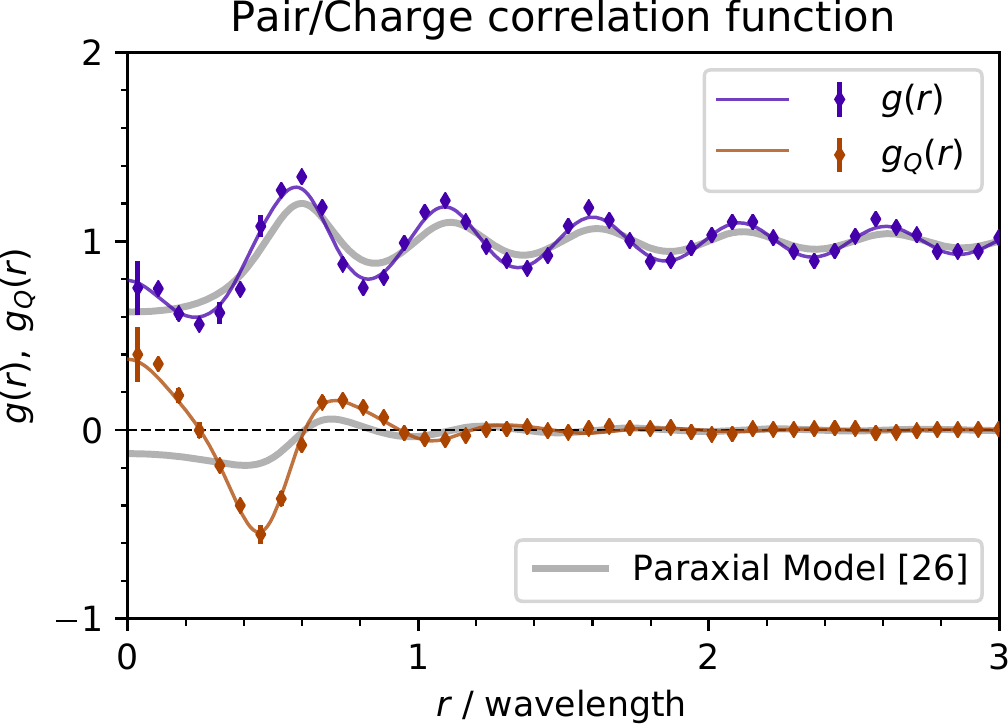}
	\caption{Pair and charge correlation function ($g$, $g_Q$) for C points in random
		vector waves. The circles are the experimental results, blue and
		yellow solid lines are our model for 2D vector fields, gray lines are the paraxial
		model \cite{Dennis2002}.}
	\label{fig:g_gQ}
\end{figure}

Figure \ref{fig:g_gQ} presents the pair correlation function for
C points in two-dimensional random light, as obtained
from our experimental data. 
With the position of each singularity known, we can compute their
pairwise distances $|\mathbf{r}_i - \mathbf{r}_j|$,
and eventually the pair correlation function
\begin{equation}
g(r) = \frac{1}{N\rho} \langle{\sum_{i \neq j} \delta ( r -|\mathbf{r}_i - \mathbf{r}_j|)}\rangle ,\label{eq:g_exp}
\end{equation}
where $N$ is the total number of singularities, $\rho$ is the average density 
of surrounding singularities and $\delta$ the Dirac function.
We compute the average and uncertainty of such a correlation function
by combining the outcome of 20 near-field measurements of the
optical random field under investigation.
In each of these maps we precisely pinpoint the location and topological
charge of approximately 6500 C points, with a spatial accuracy
which is limited by the pixel size of the experiment ($\approx 20$ nm).

$g(r)$ is not
flat, indicating that C points in
random light exhibit spatial correlation.
At first glance, this $g(r)$ seems
similar to the one of
phase singularities in scalar random
waves~\cite{Berry2000,DeAngelis2016},
and therefore also reminiscent
of that of particles in a simple liquid.
In fact,
$g(r)$ displays a damped oscillatory
behavior around unity as a function
of $r$,
with a maximum, representative of a surplus of singularities,
at approximately half a wavelength of distance.
Surprisingly, the pair correlation of C points in 2D actually increases as
$r$ approaches 0. While the zero dimensionality of optical singularities
would in principle allow for a finite
probability of having two at the same location,
an increase of $g(r)$ towards zero has never
been observed, neither for phase
singularities in scalar/vector random waves~\cite{Berry2000,DeAngelis2016},
nor for C points in a 2D slice of a 3D random field
(\cite{Dennis2002} and gray lines in Fig.~\ref{fig:g_gQ}).

To understand the unexpected behavior
at small distances and to obtain an overview of the spatial
distribution of the C points, it is useful to also consider
the charge correlation function $g_Q(r)$:
a more general expression of the pair correlation function
in which each singularity is weighted with its topological
charge~\cite{Berry2000}.
The orange data points in Fig.~\ref{fig:g_gQ} display our experimental
results for $g_Q(r)$.
The most striking observation here is
that the charge correlation function
is positive near $r=0$. This means that when
singularities are found at a close distance from each
other they most often carry
the same topological charge.
Then, at $r\approx\lambda/4$ the
charge correlation function flips sign, indicating
the beginning of a 
displacement range where two singularities are more likely
to have opposite sign.
The zero crossing roughly coincides with the
distance at which $g(r)$ exhibited
the unexpected increase towards small $r$. This
increase can therefore be attributed to
the surplus of same-sign singularities
in such a displacement range.

The reason why C points in 2D tend to rearrange
so to form closely spaced pairs with
the same topological charge
is at this stage still unclear. However, the
topological charge is not the only
intrinsic property carried by C points. More
insight could come by analyzing their behavior
with respect to light's handedness.

\subsection{C points and Light's Handedness}\label{sec:Chanddistr}
\begin{figure}[t]
	\centering
	\includegraphics[width = 8cm]{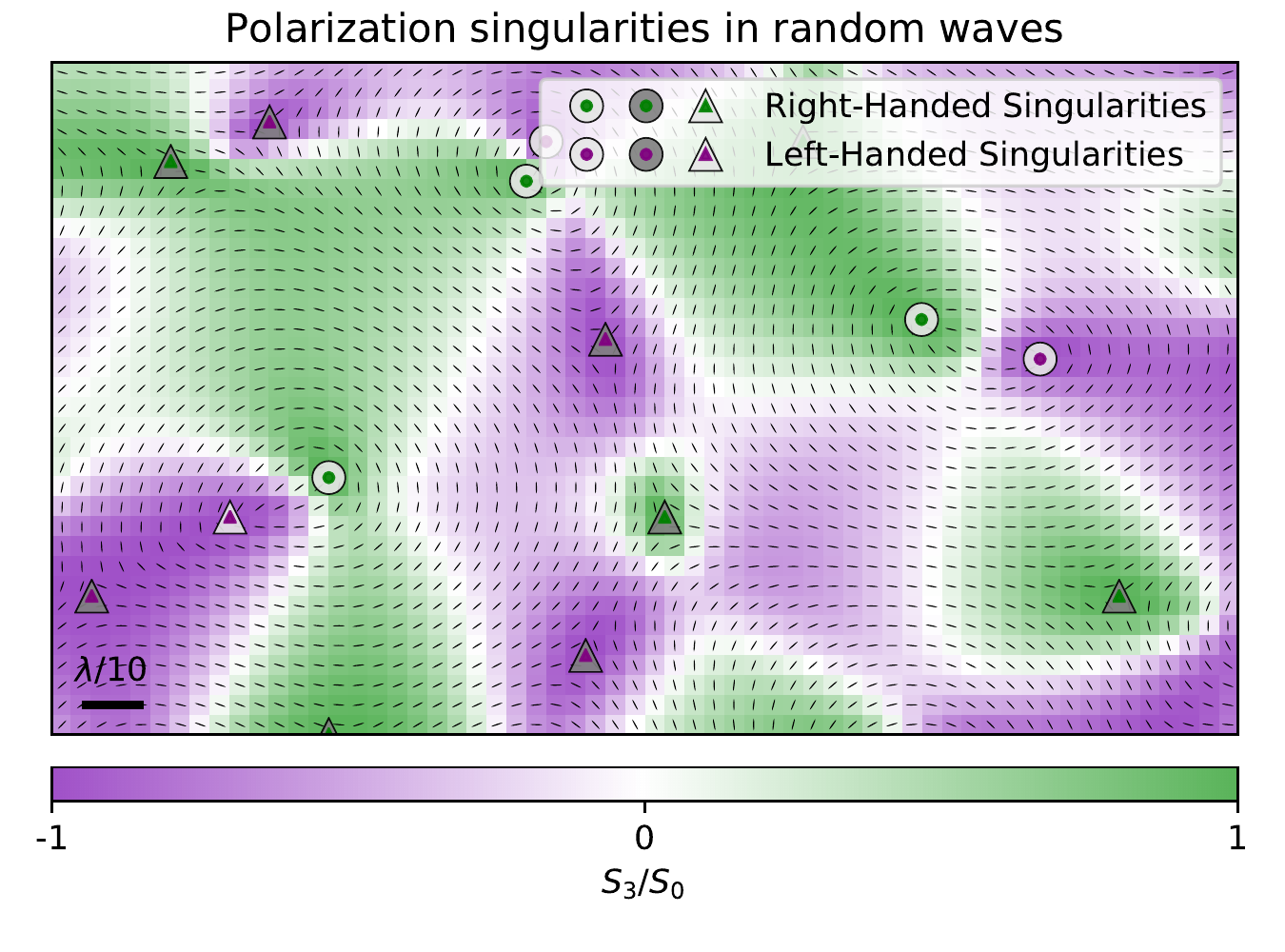}
	\caption{False-color map for the degree of circular polarization $s_3 = S_3/S_0$, as obtained from
		our experimental data. The plot corresponds to the same subsection of the measured optical random field
		displyed in Fig~\ref{fig:polsings}. The black directors indicate the orientation of the polarization ellipse.
		Circles and triangles are C points, and their filling color (purple or green) represents their handedness
		(left or right, respectively.}
	\label{fig:PS_S3}
\end{figure}

The correlation functions displayed in Fig.~\ref{fig:g_gQ}
provide an extensive description of the distribution of C points,
but still not the full picture. This is because
the information carried by C points
is not limited to their topological charge. In fact,
light's polarization is purely circular at every
C point, however it can be left- or right-handed,
independent of the topological charge.
In Fig.~\ref{fig:PS_S3} we show a spatial
map of the degree of circular polarization
$s_3 = S_3/S_0$, together with the position,
topological charge and handedness of the C points therein.
We notice how C points fall in domains of
a given handedness. Of course, $s_3$ equals exactly $+1$
or $-1$ at every
C point, with a sign which determines the handedness
of the C-point itself.
Each domain is delimited by L lines (white lines),
where polarization is purely linear ($s_3 = 0$),
and light's handedness is undetermined.
L lines have to separate C points of opposite handedness.
Contrarily, several co-handed singularities
can occur within the same
domain. Furthermore, from Fig.~\ref{fig:PS_S3} one immediately realizes
how handedness and topological charge of a C point
are not directly related, as every combination of these quantities
is possible.

\begin{figure*}[t]
	\centering
	\includegraphics[width = 17cm]{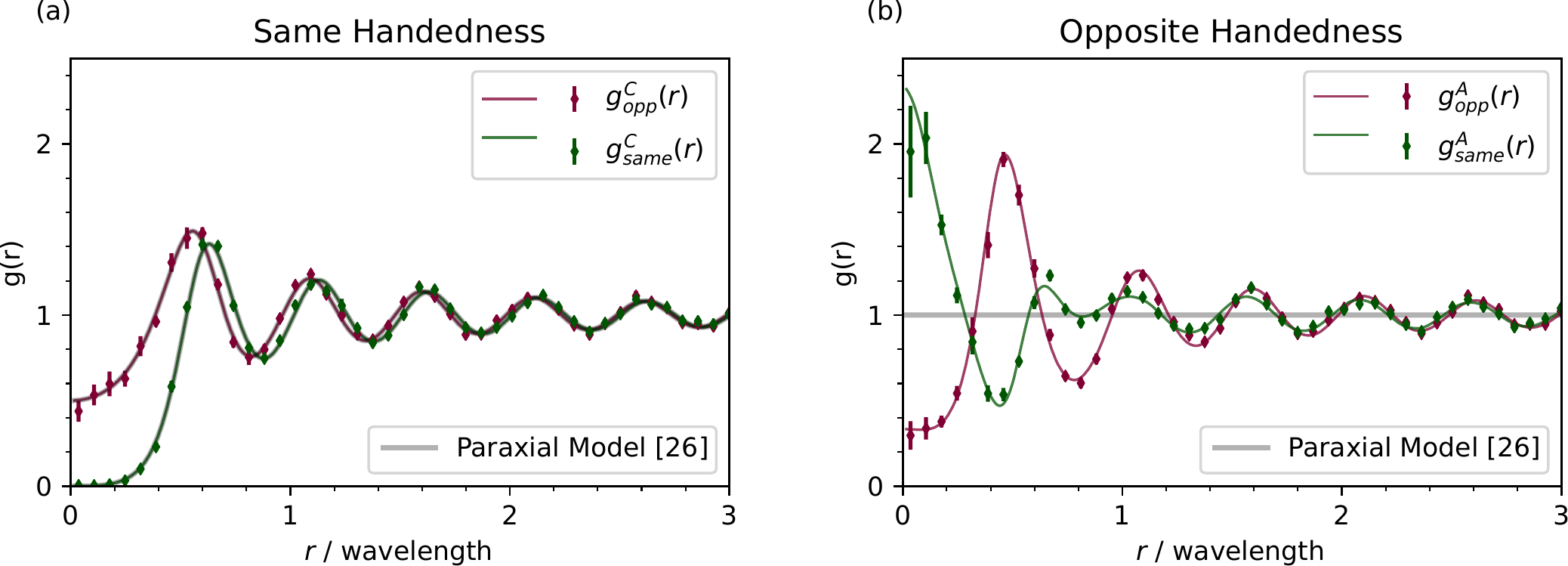}
	\caption{Pair correlation function $g(r)$ for C points with same (a) or
		opposite (b)
		handedness, and same ($g_{same}$) or opposite ($g_{opp}$) topological
		charge. Data points represent our experimental results, 
		colored solid lines are our model for isotropic 2D random field,
		solid gray lines are the 3D paraxial model~\cite{Dennis2002}. The solid
		gray lines in (a) overlap exactly with the colored solid lines.} 
	\label{fig:g_hand}
\end{figure*}

The handedness of C points provides an
additional degree of freedom to be accounted for in
their spatial distribution.
It is illuminating to include this degree of freedom in the
computation of a new set of pair correlation functions.
In general, $g(r)$ can be expressed as the
average of all the possible partial correlation
functions for C points with the same or opposite handedness and
the same or opposite topological charge:
\begin{equation}
g(r) = \frac{1}{16} \sum_{i,j} \sum_{\alpha,\beta} g_{i,j}^{\alpha,\beta} (r),
\label{eq:g_dec}
\end{equation}
where $i,j \in [+,-]$ are indices for topological charge
and $\alpha,\beta \in [l,r]$ indicate handedness. Following
the notation of Dennis \cite{Dennis2002}, Eq.~\eqref{eq:g_dec}
can be simplified with the definition of
\begin{equation}
g^C_{same} \equiv g_{i,i}^{\alpha,\alpha}\quad\mbox{and }\quad g^C_{opp } \equiv  g_{i,-i}^{\alpha,\alpha}\, ,
\end{equation}
both correspondent to co-handed singularities, for the cases
of same and opposite
toplogical charge, respectively.
Analogously, for anti-handed C points we have
\begin{equation}
g^A_{same}  \equiv  g_{i,i}^{\alpha,\bar{\alpha}}\quad\mbox{and }\quad
g^A_{opp}   \equiv  g_{i,-i}^{\alpha,\bar{\alpha}}\, .
\end{equation}
Thus, we can express Eq.~\eqref{eq:g_dec} as a function of
these four correlation functions:
\begin{equation}
g(r) = \frac{1}{4}\left[g^C_{same} + g^C_{opp} + g^A_{same} + g^A_{opp}\right].
\label{eq:g_sdec}
\end{equation}

Figure \ref{fig:g_hand} presents our experimental results for the
four pair correlation functions
of the decomposition in Eq.~\ref{eq:g_sdec},
taking both topological charge and handedness of the C points into account.
In the distribution functions depicted in
Fig.~\ref{fig:g_hand}(a) we only consider co-handed C points,
either with same (green) or opposite (purple)
topological charge. 
In this cases, the experimentally determined functions
describe the standard characteristic
properties exhibited by phase singularities in random waves.
In fact, $g^C_{same}(r\rightarrow 0) = 0$
for singularities with the same topological charge, and a monotone
decrease towards finite value at $r\rightarrow 0$ in $g^C_{opp}$.
The experimental results displayed in Fig.~\ref{fig:g_hand}(a)
perfectly match the prediction
of the model for polarization singularities in a 2D slice of
a 3D field in the paraxial regime~\cite{Dennis2002}, which is equivalent
to the model for phase singularities in scalar random waves \cite{Berry2000}.

In fact, we can interpret C points as phase singularities in
either in the left- or right-handed circular components of $\mathbf{E}$:
\begin{equation}
\psi_{l} = E_x + iE_y,\quad \psi_{r} = E_x - iE_y.
\label{eq:psiLR}
\end{equation}
This is because a phase singularity in
$\psi_l$ corresponds to a zero in $\psi_{l}$, resulting in a point where $\mathbf{E}$
has only contribution from its circular-right component $\psi_r$, i.e., a right-handed
C point. And vice versa. Therefore, the spatial distribution of co-handed C points
is exactly equivalent to that of phase singularities arising in a single
circular field component $\psi_{l/r}$, i.e., of phase singularities
in a scalar random wave field~\cite{Dennis2002}.

Our experiment confirms that also in 2D the distribution of
co-handed C points is the same as that of phase singularities in
a scalar random field. Therefore, the origin of the unusual
behavior of the global distribution
of C points must necessarily lie in anti-handed singularities.
Figure~\ref{fig:g_hand}(b) presents the correlation functions for
singularities with opposite handedness.
$g^A_{same}(r)$
reaches its maximum values at $r\approx 0$.
Singularities of opposite handedness and same topological charge
are often found at close distances from each
other, confined in an extremely subwavelength regime.
Regarding pairs of C points with opposite topological
charge, the distribution $g^A_{opp}$
exhibits a behavior that is qualitatively highly similar
to that of $g_{opp}^C$.
This creates two clearly distinct behaviors for
the four combinations of charge and handedness.
On the one hand, the impact of the handedness
of C points on their spatial correlations
seems to be only minor
for singularities with opposite topological charge, for which
we do not observe big qualitative differences between
$g_{opp}^C$ and $g_{opp}^A$
(purple data in Fig.~\ref{fig:g_hand}).
On the other hand, considering the same or opposite handedness is
crucial in the same-charge case, in which the behavior
$g_{same}^C$ and $g_{same}^A$
is evidently different, eventually with an opposite
gradient for $r\rightarrow 0$
(green data in Fig.~\ref{fig:g_hand}).

As a matter of fact, the data displayed in Fig.~\ref{fig:g_hand}(b)
offer a clear illustration of the novel behavior registered for
C points in 2D random light compared to the case
of a 2D slice of a 3D field. Especially, it clarifies
that in the 2D case C points of opposite handedness are
far from being independent, and so must be for
the left- and right-handed
field projections from which they arise.

\section{Correlation among light's vector components}

The overall spatial correlation of C points in 2D random light (Fig.~\ref{fig:g_gQ}),
and more specifically the correlation of singularities with opposite handedness [Fig.~\ref{fig:g_hand}(b)],
exhibit a number of features that
were not accounted for in a previous paraxial theory~\cite{Dennis2002}.
In that theory, an assumption was made,
consisting of the absence of any correlation between oppositely
handed C points, i.e.,
$g_{same}^A = g_{opp}^A = 1$.
This
assumption corresponds to a situation in which
 $\psi_l$ and $\psi_r$ are completely uncorrelated. 

In fact, in three-dimensions there
are no restrictions that would imply a correlation among
the circular components 
$\psi_{l}$ and $\psi_r$
of a paraxial random field. The
same holds true for a two-dimensional slice
of such a three-dimensional field~\cite{Flossmann2005}.
In this circumstance, transversality can be fulfilled
out of the plane in which
the field is observed, meaning that the vector components
of such a field can even be independently generated.
Contrarily, in a truly two-dimensional vector field transverse
propagation must be fulfilled in the same plane in which the waves
are actually propagating. 
Dismissing the third dimension while
obeying transversality
then results in a correlation among
the vector components of the field, eventually
its left- and right-handed projections.

We
will now adapt the paraxial model
of Dennis \cite{Dennis2002} in order to account
for the correlations intrinsic
to a 2D light field.
The key for explaining our results is that in
our system the electric field can be modeled as
a superposition of TE waves only.
Note that we would find completely equivalent
results considering the in-plane component
of a field composed only of TM waves \cite{Supplemental}.
A TE mode in 2D can be expressed starting from
a scalar field $H_z$:
\begin{align}\label{eq:tefield}
\left\{
\begin{aligned}
E_x & = & k_y H_z\\
E_y & = & -k_x H_z&,
\end{aligned}
\right.
\end{align}
which by default satisfies
the transverse condition.

For a random wave field, we follow Berry's hypothesis and assume
$H_z$ to be an isotropic superposition of monochromatic plane waves,
each of them with a random phase $\delta_\mathbf{k}$ \cite{Berry2000},
\begin{equation}
H_z = \sum_{|\mathbf{k}| = k_0}\exp({i\mathbf{k}\cdot\mathbf{r} +i \delta_\mathbf{k}}),
\end{equation}
where $\delta_\mathbf{k}$ is a random variable uniformly distributed in $[0,2\pi]$.
The autocorrelation of such a scalar random wave field is well known~\cite{Berry2000}:
this is a Bessel function of order zero,
\begin{equation}
	C_{zz}(\mathbf{r})  = 
	\int d\mathbf{r}_0 \,H_z^*(\mathbf{r}_0)H_z(\mathbf{r}_0+\mathbf{r})\\
	 =  J_0(k_0 r) .
	\label{eq:Cscal}
\end{equation}
The autocorrelation of $E_x$ and $E_y$ are also known~\cite{DeAngelis2016},
the main difference with $C_{zz}(\mathbf{r})$ being an anisotropic term dependent on
the orientation $\varphi$ of $\mathbf{r}$:
\begin{align}
\begin{aligned}
	C_{xx}(\mathbf{r}) & = & \frac{1}{2} \left[ J_0(k_0 r) + \cos(2\varphi)\, J_2(k_0 r) \right] &,\\
	C_{yy}(\mathbf{r}) & = & \frac{1}{2} \left[ J_0(k_0 r) - \cos(2\varphi)\, J_2(k_0 r) \right] &.
	\end{aligned}
\label{eq:Cxx}
\end{align}
Highly relevant to our study is also the cross term among $E_x$ and $E_y$,
which exhibit the following correlation:
\begin{align}
	\begin{aligned}
		C_{xy}(\mathbf{r}) 	& = &  					\int d\mathbf{r}_0 \,E_x^*(\mathbf{r}_0)E_y(\mathbf{r}_0+\mathbf{r})\\
						   	& = &\frac{1}{2}\, \sin(2\varphi)\, J_2(k_0 r) .
	\end{aligned}
\label{eq:Cxy}
\end{align}
This equation can be easily proven by carrying out the integral in Fourier space
and substituting the relations
 $E_x(\mathbf{k}) \propto \sin(\theta_{\mathbf{k}})\,\delta(|\mathbf{k}|-k_0)$
and $E_y(\mathbf{k})\propto -\cos(\theta_{\mathbf{k}})\,\delta(|\mathbf{k}|-k_0)$ \cite{DeAngelis2016}.
It is interesting to note that $E_x$ and $E_y$ only exhibit correlation
when displaced, since $C_{xy}(\mathbf{r})$ lacks the term proportional to $J_0$,
and $J_2(0) = 0$.

With these correlation functions known, and given the expression of $\psi_{l}$
and $\psi_{r}$ [Eq.~\eqref{eq:psiLR}], we have
all the ingredients to compute the correlations among the circular components
of a TE random vector field. The autocorrelation of the left-handed component is
\begin{align}\label{eq:Cll}
	\begin{aligned}
		C_{ll}(\mathbf{r}) & = &
		\int d\mathbf{r}_0 \,\psi_l^*(\mathbf{r}_0)\,\psi_l(\mathbf{r}_0+\mathbf{r})\\
		& = & C_{xx}(\mathbf{r}) + C_{yy}^*(\mathbf{r}) & = & J_0(k_0 r),
	\end{aligned}
\end{align}
and the same for $C_{rr}(\mathbf{r})$.
The result of Eq.~\eqref{eq:Cll} is also identical
to what obtained in Eq.~\eqref{eq:Cscal} for $H_z$,
proving that each separate circular component behaves as a
random scalar field.
Similarly to Eq.~\eqref{eq:Cll}, we can finally determine
the correlation among left and right circular components:
\begin{equation}\label{eq:Clr}
		C_{lr}(\mathbf{r})  =  \left[\cos(2\varphi) - i\,\sin(2\varphi)\right]\, J_2(k_0 r),\\
\end{equation}
and
\begin{equation}\label{eq:Crl}
		C_{rl}(\mathbf{r})  =  \left[\cos(2\varphi) + i\,\sin(2\varphi)\right]\, J_2(k_0 r).
\end{equation}

As elegantly explained by Berry and Dennis \cite{Berry2000},
the autocorrelation function of a complex field contains all the information needed
to retrieve the pair/charge correlation function of its phase singularities.
In the case of C points, i.e., phase singularities in the right- or left-handed field
component, also the cross-terms ($C_{rl}$ and $C_{lr}$) are necessary. 
Following the same procedure of Berry and Dennis,
we first calculate the point density of singularities in a scalar
complex field, e.g., $\psi_l \equiv \psi_l' + i\psi_l''$, which is defined as
\begin{equation}
\rho[\mathbf{u}_l] = \delta(\psi_l')\delta(\psi_l'')
\left|\frac{\partial\psi_l'}{\partial x}
\frac{\partial\psi_l''}{\partial y}-
\frac{\partial\psi_l'}{\partial y}
\frac{\partial\psi_l''}{\partial x}\right|,
\end{equation}
where $\delta$ indicates the one-dimensional Dirac's delta function, and where
for compactness we have introduced the real vector $\mathbf{u}_l = [\psi_l',\psi_l'',\partial_x \psi_l',\partial_y \psi_l',\partial_x \psi_l'',\partial_y \psi_l'']^T$. An analogous density can be defined for $\psi_r$.

The pair correlation function between
C points at two different space points $\mathbf{r}_A$ and $\mathbf{r}_B$
and with opposite handedness can now be written in a straightforward way as
\begin{equation}\label{pair_theory}
g^{A}(\mathbf{r}_B - \mathbf{r}_A) = \frac{\langle\, \rho[\mathbf{u}_l(\mathbf{r}_A)]\,\rho[\mathbf{u}_r(\mathbf{r}_B)] \,\rangle}
{\langle\, \rho[\mathbf{u}_l(\mathbf{r}_A)]\,\rangle\langle\,\rho[\mathbf{u}_r(\mathbf{r}_B)]\, \rangle}.
\end{equation}
In this equation, the notation $\langle\, f[\mathbf{u}_l(\mathbf{r}_A),\mathbf{u}_r(\mathbf{r}_B)]\,\rangle$ indicates the statistical average of a generic $f$, functional of the field components and of their derivatives at different points in space.
Introducing the combined vector $\mathbf{u} = [\mathbf{u}_l(\mathbf{r}_A),\mathbf{u}_r(\mathbf{r}_B)]^T$, the average can be explicitly written in the form:
\begin{equation}
\langle f[\mathbf{u}] \rangle = \frac{1}{(2\pi)^{D/2}\sqrt{\det M}}\int d^D\mathbf{u}\:f[\mathbf{u}] \exp(-\tfrac{1}{2}\mathbf{u}^T M^{-1} \mathbf{u}),
\end{equation}
where $D$ is the dimension of the vector $\mathbf{u}$ and $M$ is the matrix
of the correlations between the various components of $\mathbf{u}$,
i.e., $M_{ij} = \langle u_i u_j \rangle$.
These elements correspond to the correlations
between the different components of the left- and
right-handed fields that we have summarized above, and their
spatial derivatives.
Similar expressions for different combinations of the fields $\psi_l$
and $\psi_r$ and for specific choices of the charge of the singularities
can be obtained from Eq.~\eqref{pair_theory} with intuitive modifications.

In some particular cases \cite{Berry2000,Dennis2002,DeAngelis2016},
it is possible to derive a closed analytical expression for averages
of the form in Eq.~\eqref{pair_theory} by reducing the integrand to a quadratic form
and integrating with standard mathematical techniques \cite{Li2009}.
However, the specific form of the correlation matrix in our model does
not lend itself easily to applying the formalism of Ref.~\cite{Li2009}.
This is due to the additional correlations between the real and
imaginary parts of the field components, corresponding to the imaginary
terms in $C_{lr}$ and $C_{rl}$ [Eqs.~\eqref{eq:Clr} and \eqref{eq:Crl}].
Nevertheless, the average in Eq.~\eqref{pair_theory} is particularly suited
to numerical integration with Monte Carlo techniques \cite{Press2007}.
We therefore calculated the pair correlation functions of C points
and polarization vortices in two steps.
Firstly, we perform analytically the integral
over the terms containing the Dirac's delta functions in the integrand of
Eq.~\eqref{pair_theory}. Subsequently,
we carry out numerically the integration over the remaining variables,
using the multidimensional Monte Carlo method \cite{Press2007}.

We plot the theoretical expectations for the pair/charge correlation
functions in direct comparison with the experimental data. In Fig.~\ref{fig:g_gQ}
we show the pair and charge correlation function for C points in 2D random vector waves
and in Fig.~\ref{fig:g_hand} the pair correlation functions
for C points with the same or opposite handedness, respectively. For each of these
curves we find an excellent agreement with the experiment.
In particular, the pair correlation
functions displayed in Fig.~\ref{fig:g_hand}(b) for C points with opposite handedness represent the major novelty introduced by the model for 2D light.
Among these functions, $g_{same}^A$ exhibits a
behavior which is extremely unusual for
pair correlations of this kind. Although 
this behavior is perfectly consistent with the
experimental observation, it might conceal further
interesting properties of random light confined in 2D.

\section{Conclusions}

In this work we investigated the spatial correlation of C points in 2D random light. We
compared it to existing theory and experiments for 2D slices through a 3D random field in the paraxial regime. We demonstrated that confining the optical field to propagate in two dimensions induces severe changes in the spatial distribution of its C points. The shortage of degrees of freedom caused by the removal of one dimension results in a correlation among the vector components of the 2D light field. In the circular basis, this results in a correlation among the oppositely-handed optical-spin components of light. One of the key consequences was the observation that the chance of finding C points with same topological charge actually increases as their mutual distance goes to zero. This is an unusual finding for dislocations of any kind. We quantify the correlation between left- and right-handed spin for the case of a TE field and incorporate it in a newly developed theoretical model. Our results are general for in-plane fields, including those of a TM mode as well. The outcome of the 2D model is found to be in perfect agreement with our experimental results. Given the unusual properties of the ensemble of C points in 2D random vector waves, our findings may trigger a re-evaluation of concepts which are considered pillars of singular optics and topological defects, i.e. the sign principle \cite{Freund1994} and topological screening \cite{Freund1998}. The behavior at short distances might lead to more unexplored features such as polarization vortices and higher-order singularities.

\section*{Acknowldegments}
We thank Andrea Di Falco for fabricating the chaotic
cavity used in the near-field experiments and
Thomas Bauer for useful discussions.
This work is part of the research program of
the Netherlands Organization for Scientific Research
(NWO). The authors acknowledge funding from the
European Research Council (ERC Advanced Grant
No. 340438-CONSTANS). F. A. acknowledges support
from the Marie Skłodowska-Curie individual fellowship
BISTRO-LIGHT (Grant No. 748950).

%

\end{document}